\def\year{2020}\relax
\documentclass[letterpaper]{article} 
\usepackage{aaai20}  
\usepackage{times}  
\usepackage{helvet} 
\usepackage{courier}  
\usepackage[hyphens]{url}  
\usepackage{graphicx} 
\usepackage{graphicx}  
\usepackage{verbatim}

\usepackage{amsmath}
\DeclareMathOperator*{\argmax}{argmax}
\usepackage{adjustbox}
\usepackage{float}
\title{FPETS : Fully Parallel End-to-End Text-to-Speech System}

\author{ \Large \textbf{Dabiao Ma$^*$ \textsuperscript{\rm 1}, Zhiba Su$^*$ \textsuperscript{\rm 1}, Wenxuan Wang$^*$ \textsuperscript{\rm 2}, Yuhao Lu$^{\dag}$\textsuperscript{\rm 1}}\\ 
\textsuperscript{\rm 1} Turing Robot Co.,Ltd. Beijing, China\\
\{madabiao, suzhiba, luyuhao\}@uzoo.cn \\
\textsuperscript{\rm 2} The Chinese University of Hong Kong, Shenzhen. Guangdong, China\\
wenxuanwang1@link.cuhk.edu.cn \\
}

\begin{document}

\maketitle

\footnotetext[1]{Dabiao Ma, Zhiba Su and Wenxuan Wang have equal contributions. Yuhao Lu is the corresponding author.}
\footnotetext[2]{Codes and demos will be released at \url{https://github.com/suzhiba/Full-parallel_100x_real_time_End2EndTTS}}

\begin{abstract}
End-to-end Text-to-speech (TTS) system can greatly improve the quality of synthesised speech. But it usually suffers form high time latency due to its auto-regressive structure. And the synthesised speech may also suffer from some error modes, e.g. repeated words, mispronunciations, and skipped words. In this paper, we propose a novel non-autoregressive, fully parallel end-to-end TTS system (FPETS). It utilizes a new alignment model and the recently proposed U-shape convolutional structure, UFANS. Different from RNN, UFANS can capture long term information in a fully parallel manner. Trainable position encoding and two-step training strategy are used for learning better alignments. Experimental results show FPETS utilizes the power of parallel computation and reaches a significant speed up of inference compared with state-of-the-art end-to-end TTS systems. More specifically, FPETS is 600X faster than Tacotron2, 50X faster than DCTTS and 10X faster than Deep Voice3. And FPETS can generates audios with equal or better quality and fewer errors comparing with other system. As far as we know, FPETS is the first end-to-end TTS system which is fully parallel.
\end{abstract}

\section{Introduction}

TTS systems aim to generate human-like speeches from texts. End-to-end TTS system is a type of system that can be trained on (text,audio) pairs without phoneme duration annotation\cite{Tacotron}. It usually contains $2$ components, an acoustic model and a vocoder. Acoustic model predicts acoustic intermediate features from texts. And vocoder, e.g. Griffin-Lim \cite{Griffin_lim}, WORLD \cite{Merlin}, WaveNet \cite{WaveNet}, synthesizes speeches with generated acoustic features.

The advantages of end-to-end TTS system are threefold: 1) reducing manual annotation cost and being able to utilize raw data, 2) preventing the error propagation between different components, 3) reducing the need of feature engineering. However, without the annotation of duration information, end-to-end TTS systems have to learn the alignment between text and audio frame.

 Most competitive end-to-end TTS systems have an encoder-decoder  structure  with attention  mechanism, which is significantly helpful for alignment learning. Tacotron \cite{Tacotron} uses an autoregressive attention \cite{attention} structure to predict alignment, and uses  CNNs and GRU \cite{GRU}  as encoder and decoder, respectively. Tacotron2\cite{shen2018natural}, which is a
combination of the modified Tacotron system and WaveNet, also use an autoregressive attention. However, the autoregressive structure greatly limits the inference speed in the context of parallel computation. Deep voice 3 \cite{dp3}  replaces RNNs with CNNs to speed up training and inference. DCTTS \cite{DCTTS} greatly speeds up the training of attention module by introducing guided attention. But Deep Voice 3 and DCTTS  still have autoregressive structure. And those models also suffer from serious error modes e.g. repeated words, mispronunciations, or skipped words \cite{dp3}. 

Low time latency is required in real world application. Autoregressive  structures, however, greatly limit the inference speed in the context of parallel computation. \cite{dp3} claims that it is hard to learn alignment without a autoregressive structure. So the question is how to design a non-autoregressive structure that can perfectly determine alignment?

In this paper, we propose a novel fully parallel end-to-end TTS system (FPETS). Given input phonemes, our model can predict all acoustic frames simultaneously rather than autoregressively. Specifically, we follow the commonly used encoder-decoder structure with attention mechanism for alignment. But we replace autoregressive structures with a recent proposed U-shaped convolutional structure (UFANS)\cite{UFANS}, which can be fully parallel  and has stronger representation ability. Our fully parallel alignment structure inference alignment relationship between all phonemes and audio frames at once. Our novel trainable position encoding method can utilize position information better and two-step training strategy improves the alignment quality. 

Experimental results show FPETS utilizes the power of parallel computation and reaches a significant speed up of inference compared with state-of-the-art end-to-end TTS systems. More specifically, FPETS is 600X faster than Tacotron2, 50X faster than DCTTS and 10X faster than Deep Voice3. And FPETS can generates audios with equal or better quality and fewer errors comparing with other system. As far as we know, FPETS is the first end-to-end TTS system which is fully parallel.

\section{Model Architecture}

\begin{figure}[t]
    \centering

    \includegraphics[width=3.2in]{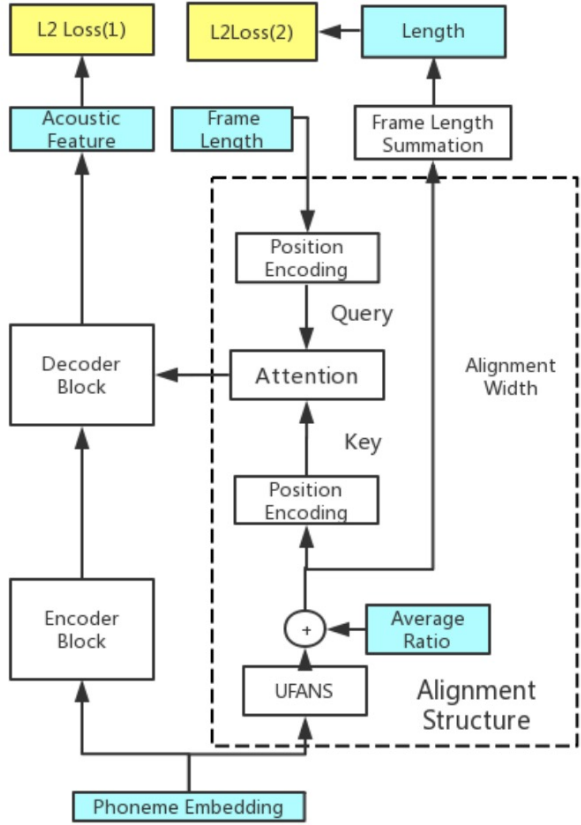}
    \caption{ Model architecture. The light blue blocks are input/output flow.}
\label{fig : overall system}
\end{figure}

Most competitive end-to-end TTS systems have an encoder-decoder structure with attention mechanism\cite{Tacotron} \cite{dp3}. Following this overall architecture, our model consists of three parts, shown in Fig.\ref{fig : overall system}. The encoder converts phonemes into hidden states that are sent to decoder; The alignment module determines the alignment width of each phoneme, from which the number of frames that attend on that phoneme can be induced; The decoder receives alignment information and converts the encoder hidden states into acoustic features. 

\subsection{Encoder}

The encoder encodes phonemes into hidden states. It consists of 1 embedding layer , 1 dense layer, 3 convolutional layers, and a final dense layer. Some of TTS systems\cite{Li2018CloseTH} use self attention network as encoder. But we find that it dose not make significant difference, both in loss value and MOS.

Alignment module determines the mapping from phonemes to  acoustic features. We discard autoregressive structure, which is widely used in other alignment modules\cite{dp3}\cite{Tacotron}\cite{shen2018natural}, for time latency issue.  Our novel alignment module consists of 1 embedding layer, 1 UFANS \cite{UFANS} structure, trainable position encoding and several matrix multiplications, as depicted in Fig.\ref{fig : overall system}. 

\subsubsection{Fully parallel UFANS structure}
\begin{figure*}[h]
    \centering
    \includegraphics[width=6in]{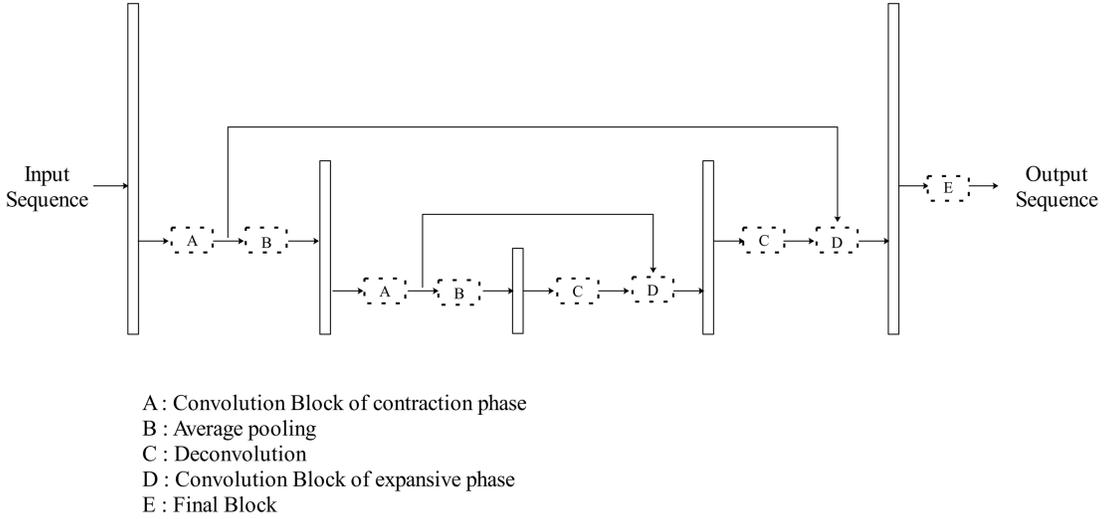}
    \caption{ UFANS Model Architecture}
\label{fig : UFANS}
\end{figure*}

UFANS is a modified version of U-Net for TTS task aiming to speed up inference. The structure is shown in Fig.\ref{fig : UFANS}.In alignment structure, UFANS is used to predict alignment width, which is similar to phoneme duration.Those pooling and up-sampling operation along the spatial dimension make the receptive field increases exponentially and high way connection enables the combination of different scales features.

For each phoneme $i$, we define the 'Alignment Width' $r_i$ which represents its relationship with frame numbers. Suppose the number of phonemes in an utterance is $ N $, and UFANS outputs a sequence of $ N $ scalars : $ [r_0, r_1, ..., r_{N - 1}] $;

Then we relate the alignment width $ r_i $ to the acoustic frame index $j$. The intuition is that the acoustic frame with index $j$= $ \sum_{k = 0}^{i - 1} r_k + \frac{1}{2} r_i $ should be the one that attends most on $ i $-th phoneme. And we need a structure that satisfies the intuition.

\subsubsection{Position Encoding Function}

Position encoding \cite{STS} is a method to embed a sequence of absolute positions into a sequence of vectors. Sine and cosine position encoding has two very important properties that make it suitable for position encoding. In brief, function $ g(x) =  \sum_f cos (\frac{x - s}{f}) $ has a heavy tail that enables one acoustic frame to receive phoneme information very far away; The gradient function  $ |\dot{g}(s)| = |\sum_f sin (\frac{x - s}{f})| $ is insensitive to the term $ x - s $. We give a more detailed illustration in Appendix.

\subsubsection{Trainable Position Encoding}

Some end-to-end TTS system, like deep voice 3 and Tacotron2, use sine and cosine functions of different frequencies and add those position encoding vectors to input embedding. But they both take position encoding as a supplement to help the training of attention module and the position encoding vectors remain constant. We propose a trainable position encoding, which is better than absolute position encoding in getting position information.
\\ \\
We define the absolute alignment position $ s_i $ of $ i $-th phoneme as :
\begin{equation}
s_i = \sum_{k=0}^{i - 1} r_k + \frac{1}{2}r_i, i = 0, ..., T_p - 1, r_{-1} = 0
\end{equation}
Now choose $ L $ float numbers log uniformly from range $ [1.0, 10000.0] $ and get a sequence of frequencies $ [f_0, ..., f_{L - 1}]$. For $ i $-th phoneme, the position encoding vector $ vp_i $ of this phoneme is defined as :
\begin{equation}
\begin{split}
&vp_i = [vp_{i, sin}, vp_{i, cos}], \\
&[vp_{i, sin}]_k = sin (\frac{s_i}{f_k}), \\
&[vp_{i, cos}]_k = cos (\frac{s_i}{f_k}), k = 0, ..., L - 1 \\
\end{split}
\end{equation}
Concatenating $ vp_i, i = 0, ..., T_p - 1 $ together, we get a matrix $ P $ that represents position information of all the phonemes, denoted as 'Key', see Fig.\ref{fig : overall system} :
\begin{equation}
P = [vp_0^T, ..., vp_{T_p - 1}^T]
\end{equation}
And similarly, for the $ j $-th frame of the acoustic feature, the position encoding vector $ va_j $ is defined as :
\begin{equation}
\begin{split}
&va_j = [va_{j, sin}, va_{j, cos}], \\
&[va_{j, sin}]_k = sin (\frac{j}{f_k}), \\
&[va_{i, cos}]_k = cos (\frac{j}{f_k}), k = 0, ..., L - 1
\end{split}
\end{equation}
Concatenating all the vectors, we get the matrix $ F $ that represents position information of all the acoustic frames, denoted as 'Query', see Fig.\ref{fig : overall system}:
\begin{equation}
F = [va_0^T, ..., va_{T_a - 1}^T]
\end{equation}
And now define the attention matrix $ A $ as :
\begin{equation}
\begin{split}
&A = FP^T, A_{ji} = vp_iva_j^T, \\
&i = 0, ..., T_p - 1, j = 0, ..., T_a - 1\\
\end{split}
\end{equation}
That is, the attention of $ j $-th frame on $ i $-th phoneme is proportional to the inner product of their encoding vectors. This inner product can be rewritten as :
\begin{equation}
\begin{split}
vp_iva_j^T &= \sum_f(cos (\frac{s_i}{f}) cos (\frac{j}{f}) + sin (\frac{s_i}{f}) sin (\frac{j}{f})) \\
&= \sum_f cos (\frac{s_i}{f} - \frac{j}{f})
\end{split}
\end{equation}
It is clear when $ j = s_i $, the $ j $-th frame is the one that attends most on $ i $-th phoneme.
The normalized attention matrix $ \hat{A} $ is :
\begin{equation}
\hat{A}, \hat{A}_{ji} = \frac{A_{ji}}{\sum_iA_{ji}}
\end{equation}
Now $ \hat{A}_{ji} $ represents how much $ j $-th frame attends on $ i $-th phoneme. 

Then we use argmax to build new attention matrix $ \widetilde{A} $: 
\begin{equation}
  \widetilde{A}_{ji} =
  \begin{cases}
  1 & \text{if}\ i = \argmax \limits_{k \in [0, ..., T_p - 1]} A_{jk} \\
  0 & \text{otherwise }
  \end{cases}
\end{equation}

Now define the number of frames that attend more on $ i $-th phoneme than any other phoneme to be its attention width $ w_i $. From the definition of attention width, $ \widetilde{A} $ is actually a matrix representing attention width $ w_i $. The alignment width $ r_i $ and $ w_i $ are different but related.

For two adjacent absolute alignment positions $ s_i $ and $ s_{i + 1} $, consider the two functions:

$$ g_1 (x) = \sum_f cos (\frac{x - s_i}{f}) ,  g_2 (x) = \sum_f cos (\frac{x - s_{i + 1}}{f}) $$

The values of the two functions only depend on the relative position of $ x $ to $ s_i$ and $s_{i + 1} $. It is known function $g_1$ decreases when $ x $ moves away from $ s_i $ (locally, but it is sufficient here). So we have: 

$$\begin{cases}
 g_1 (x) > g_2 (x)  & \text{when } x \in [s_i, \frac{1}{2}(s_i + s_{i + 1}))\\
 g_1 (x) < g_2 (x)  & \text{when } x \in (\frac{1}{2}(s_i + s_{i + 1}), s_{i + 1}]
\end{cases}$$

Thus $ x = \frac{1}{2}(s_i + s_{i + 1}) $ is the right attention boundary of phoneme $ i $, similarly the left attention boundary is $ x = \frac{1}{2}(s_{i - 1} + s_i) $. It can be deduced that :
\begin{align}
w_i &= \frac{1}{2}(s_i + s_{i + 1}) - \frac{1}{2}(s_{i - 1} + s_i) \\
&= \frac{1}{4}(r_{i-1} + r_{i+1} + 2r_i) \\
i &= 0, ..., T_p - 1, r_{-1} = r_0, r_{T_p} = r_{T_P - 1}
\end{align}
which means attention width and alignment width can be linearly transformed to each other. And it is further deduced that : 
\begin{equation}
\sum_{k=0}^{T_p - 1}r_k = \sum_{k=0}^{T_p - 1}w_k = T_a
\end{equation}

\begin{figure}[t]
    \centering
    \includegraphics[width=3in]{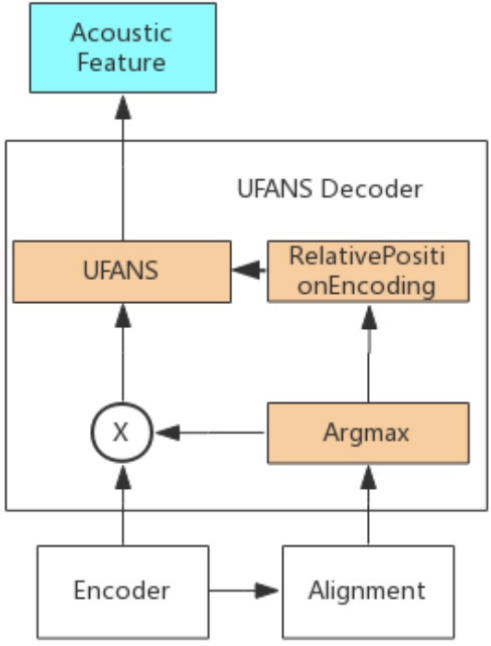}
    \caption{UFANS Decoder}
\label{fig : UDecoder}
\end{figure}

\subsection{UFANS Decoder}
 The decoder receives alignment information and converts the encoded phonemes information to acoustic features, see Figure \ref{fig : UDecoder}. Relative position is the distance between the phoneme and previous phoneme. Our model use it to enhance position relationship. Following \cite{UFANS}, we use UFANS as our decoder. The huge receptive field enables to capture long-time information dependency and the high-way skip connection structure enables the combination of different level of features. It generates good quality acoustic features in a fully parallel manner.

\subsection{Training Strategy}

We use Acoustic Loss, denoted as $ LOSS_{acou}$, to evaluate the quality of generated acoustic features, which is $ L_2 $ norm between predicted acoustic features and ground truth features.

In order to train a better alignment model, we propose a two-stage training strategy. Our model focus more on alignment learning in stage 1. In stage 2 we fix the alignment module and train the whole system. 

\subsubsection{Stage 1 :Alignment Learning}
In order to enhance the quality of alignment learning, we use convolutional decoder and design an alignment loss.\\
Convolutional Decoder: UFANS has stronger representation ability than vanilla CNN. But the learning of alignment will be greatly disturbed if using UFANS as decoder. The experimental evidences and analysis are shown in the next section. 

So we replace UFANS decoder with a convolutional decoder. The convolutional decoder consists of several convolution layers with gated activation \cite{GatedActivation}, several Dropout \cite{Dropout} operations and one dense layer.\\
Alignment Loss: We define an Alignment Loss, denoted as $ LOSS_{align} $, based on the fact that the summation of alignment width should be equal or close to the frame length of acoustic features. We relax this restriction by using a threshold $ \gamma $ :
\begin{equation}
    LOSS_{align}=
    \begin{cases}
      \gamma, & \text{if}\  |\sum_{k=0}^{T_p - 1}r_k - T_a|\\ 
                         &    < \gamma \\
      |\sum_{k=0}^{T_p - 1}r_k - T_a|, & \text{otherwise}
    \end{cases}
  \end{equation}
The final loss $ LOSS $ is a weighted sum of $ LOSS_{acou} $ and $ LOSS_{align} $ : 
\begin{equation}
LOSS = LOSS_{acou} + \sigma LOSS_{align}
\end{equation}
We choose $0.02$ as alignment loss weight based on grid search from $0.005$ to $0.3$.

\subsubsection{Stage 2 : Overall Training}

In stage 2, we fix the well-trained alignment module and use UFANS as decoder to train the overall end-to-end system. Only Acoustic Loss is used as objective function in this stage.
\section{Experiments and Results}

\subsection{Dataset}
LJ speech\cite{LJSpeech} is a public speech dataset consisting of 13100 pairs of text and 22050 HZ audio clips. The clips vary from 1 to 10 seconds and the total length is about 24 hours. Phoneme-based textual features are given. Two kinds of acoustic features are extracted. One is based on WORLD vocoder that uses mel-frequency cepstral coefficients(MFCCs). The other is linear-scale log magnitude spectrograms and mel-band spectrograms that can be feed into Griffin-Lim algorithm or a trained WaveNet vocoder.

The WORLD vocoder uses 60 dimensional mel-frequency cepstral coefficients, 2 dimensional band aperiodicity, 1 dimensional logarithmic fundamental frequency, their delta, delta-delta dynamic features and 1 dimensional voice/unvoiced feature. It is 190 dimensions in total. The WORLD vocoder based feature uses FFT window size 2048 and has a frame time 5 ms. 

The spectrograms are obtained with FFT size 2048 and hop size 275. The dimensions of linear-scale log magnitude spectrograms and mel-band spectrograms are 1025 and 80.

\subsection{Implementation Details}

Hyperparameters of our model are showed in Table \ref{tab : Hyper-parameter}. Tacotron2, DCTTS and Deep Voice3 are used as baseline . The model configurations are shown in Appendix. Adam are used as optimizer with $\beta_1 = 0.9$, $\beta_2 = 0.98$, $\epsilon = 1e-4$. Each model is trained 300k steps.

All the experiments are done on 4 GTX 1080Ti GPUs, with batch size of 32 sentences on each GPU.

\begin{table}
\centering
\hspace{4pt}

\caption{Hyper-Parameter}

\begin{tabular}{l|c}
\hline
Structure & value  \\
\hline
Encoder/DNN Layers &  1  \\

Encoder/CNN Layers &  3  \\

Encoder/CNN Kernel &  3  \\

Encoder/CNN Filter Size &  1024  \\

Encoder/Final DNN Layers &  1  \\
\hline
Alignment/UFANS layers& 4 \\

Alignment/UFANS hidden& 512 \\

Alignment/UFANS Kernel& 3 \\

Alignment/UFANS Filter Size& 1024 \\
\hline
CNN Decoder/CNN Layers& 3 \\

CNN Decoder/CNN Kernel& 3 \\

CNN Decoder/CNN Filter Size& 1024 \\
\hline
UFANS Decoder/UFANS layers& 6 \\

UFANS Decoder/UFANS hidden& 512 \\

UFANS Decoder/UFANS Kernel& 3 \\

UFANS Decoder/UFANS Filter Size& 1024 \\
\hline
Droupout& 0.15 \\
\hline
\end{tabular}
\label{tab : Hyper-parameter}
\end{table}

\subsection{Main Results}
We aim to design a TTS system that can synthesis speech quickly, high quality and with fewer errors.  So we compare our FPUTS with baseline on inference speed, MOS and error modes.

\subsubsection{Inference Speed}
The inference speed evaluates time latency of synthesizing a one-second speech, which includes data transfer from main memory to GPU global memory, GPU calculations and data transfer back to main memory.  As is shown in Table~\ref{tab : inf speed}, our FPETS model is able to greatly take advantage of parallel computations and is significantly faster than other systems.

\subsubsection{MOS}
Harvard Sentences List 1 and List 2 are used to evaluate the mean opinion score (MOS) of a system. The synthesized audios are evaluated on Amazon Mechanical Turk using crowdMOS method \cite{crowdMOS}. The score ranges from 1 (Bad) to 5 (Excellent). As is shown in Table~\ref{tab : mos}, Our FPETS is no worse than other end-to-end system. The MOS of WaveNet-based audios are lower than expected since background noise exists in these audios.

\subsubsection{Robustness Analysis}
 Attention-based neural TTS systems may run into several error modes that can reduce synthesis quality. For example, repetition means repeated pronunciation of one or more phonemes, mispronunciation means wrong pronunciation of one or more phonemes and skip word means one or more phonemes are skipped.

In order to track the occurrence of attention errors, 100 sentences are randomly selected from Los Angeles Times, Washington Post and some fairy tales. As is shown in Table~\ref{tab : err}, Our FPETS system is more robust than other systems.

\begin{table}[t]
\normalsize
\caption{Inference speed comparison}
\centering
\resizebox{.95\columnwidth}{!}{
\begin{tabular}{l|c|c}
\hline
Method& Autoregressive & Inference speed (ms)  \\
\hline
Tacotron2 &Yes  & 6157.3 \\

 DCTTS & Yes & 494.3 \\

Deep Voice 3 & Yes & 105.4 \\

\textbf{FPETS} & \textbf{No} &  \textbf{9.9} \\
\hline
\end{tabular}
}
\label{tab : inf speed}

\hspace{6pt}
\centering
\caption{MOS results comparison}
\begin{tabular}{l|c|c}
\hline
Method & Vocoder & MOS \\
\hline
Tacotron2 &  Griffin Lim & $ 3.51 \pm 0.070 $  \\

DCTTS &  Griffin Lim & $ 3.55 \pm 0.107 $ \\

Deep Voice 3 &  Griffin Lim & $ 2.79 \pm 0.096 $ \\

\textbf{FPETS}  & \textbf{Griffin Lim} & \textbf{$ 3.65 \pm 0.082 $} \\
\hline
Tacotron2  & WaveNet & $ 3.04 \pm 0.103 $  \\

DCTTS  & WaveNet & $ 3.43 \pm 0.109 $ \\

FPETS  & WaveNet & $ 3.27 \pm 0.108 $ \\
\hline
\textbf{FPETS} & \textbf{WORLD} & \textbf{$ 3.81 \pm 0.122 $}\\
\hline
\end{tabular}
\label{tab : mos}
\hspace{6pt}

\caption{Robustness Comparison}
\centering
\begin{adjustbox}{width=0.45\textwidth}
\begin{tabular}{l|c|c|c}
\hline
& Repeats & Mispronunciation & Skip \\
\hline
Tacotron2 & 2 &  5 & 4 \\

DCTTS & 2 &  10 & 1 \\

Deep Voice 3 & 1&  5 & 3 \\

\textbf{FPETS} & \textbf{1}  & \textbf{2} & \textbf{1}\\
\hline
\end{tabular}
\end{adjustbox}

\label{tab : err}
\end{table}

\begin{table*}[h]
\centering
\caption{A case study about phoneme-level comparison of alignment quality. Real duration and the predicted duration by our alignment method, using Gaussian as position encoding function, using fixed position encoding, using UFANS as decoder are shown.}
\begin{adjustbox}{width=\textwidth}
\begin{tabular}{l|c|c|c|c|c|c|c|c|c|c|c|c|c|c|c|c}
\hline
& P & R & AY & ER & T & UW & N & OW & V & EH & M & B & ER & T & W & EH  \\
\hline
real & 5.35 & 7.28 & 15.48 & 13.43 & 4.96 & 3.44 & 3.36 & 5.44 & 4.72 & 7.20 & 4.56 & 1.92 & 7.12 & 5.36 & 3.36 & 3.84 \\

resynth & 3.55 & 7.97 & 13.28 & 11.37 & 4.88 & 4.00 & 6.19 & 5.27 & 5.46 & 6.39 & 3.56 & 2.08 & 6.13 & 5.69 & 4.34 & 3.03\\

resynth-Gauss & 6.31 & 6.03 & 5.78 & 6.11 & 6.59 & 6.73 & 6.74 & 6.76 & 6.75 & 6.75 & 6.77 & 6.80 & 6.84 & 6.82 & 6.79 & 6.78 \\

resynth-fixenc & 7.41 & 7.35 & 11.40 & 10.46 & 4.04 & 4.60 & 2.95 & 6.41 & 4.30 & 7.86 & 5.26 & 2.45 & 9.21 & 6.77 & 3.90 & 2.85\\

resynth-UFANS & 4.08 & 8.09 & 9.41 & 8.45 & 6.90 & 5.70 & 5.21 & 5.71 & 5.98 & 5.29 & 4.87 & 5.20 & 5.43 & 5.14 & 5.10 & 5.37\\

\hline
& IY & T & UW & N & AY & N & T & IY & N & S & IH & K & S & T & IY & TH \\
\hline
real & 10.80 & 9.76 & 9.76 & 6.80 & 6.08 & 6.16 & 7.28 & 5.28 & 5.36 & 6.56 & 6.16 & 4.08 & 3.52 & 6.32 & 9.36 & 9.76 \\

resynth & 10.89 & 11.26 & 9.69 & 7.72 & 5.33 & 6.55 & 7.30 & 5.90 & 5.81 & 5.43 & 5.11 & 4.33 & 3.57 & 6.81 & 10.57 & 11.54\\

resynth-Gauss & 6.78 & 6.79 & 6.77 & 6.75 & 6.76 & 6.74 & 6.72 & 6.74 & 6.76 & 6.80 & 6.84 & 6.82 & 6.79 & 6.78 & 6.77 & 6.81\\

resynth-fixenc & 12.05 & 9.21 & 8.26 & 5.76 & 6.90 & 7.63 & 6.47 & 3.20 & 4.74 & 4.11 & 5.85 & 2.97 & 4.01 & 5.29 & 11.26 & 10.60  \\

resynth-UFANS & 7.14 & 7.38 & 8.96 & 8.20 & 5.39 & 5.54 & 7.31 & 6.34 & 5.56 & 6.42 & 5.84 & 4.76 & 5.62 & 7.61 & 8.26 & 8.17\\
\hline

\end{tabular}
\end{adjustbox}

\label{tab : real and syn comp}
\end{table*}

\subsection{Alignment Learning Analysis}

Alignment learning is essential for end-to-end TTS system which greatly affects the quality of generated audios. So we further discuss the factors that can affect the alignment quality.

\begin{figure}[H]
  \centering
  \includegraphics[width=3in]{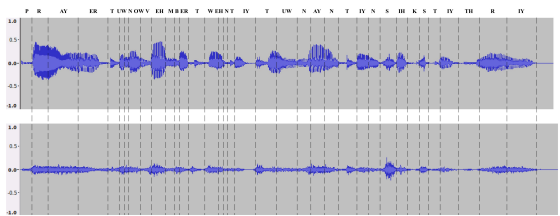}
  \caption{The upper is real audio of 'LJ048-0033', the lower is the re-synthesized audio from alignment learning model.\\
           text : prior to November twenty two nineteen sixty three \\
           phoneme : P R AY ER T UW N OW V EH M B ER T W EH N T IY T UW N AY N T IY N S IH K S T IY TH R IY}
\label{fig : real and syn}
\end{figure}

100 audios are randomly selected from training data, denoted as origin audios. Their utterances are fed to our system  to generate audios, denoted as re-synthesized audios. The method to evaluate the alignment quality is objectively computing the difference of the phoneme duration between origin audios and their corresponding re-synthesized audios. The phoneme durations are obtained by hand. Figure~\ref{fig : real and syn} is the labeled phonemes of audio 'LJ048-0033'. Here only results with mel-band spectrograms using Griffin-Lim algorithm are shown. For MFCCs, results are similar.

We compare alignment quality between different alignment model configurations. Table~\ref{tab : ave diff} shows the overall results on 100 audios. Table~\ref{tab : real and syn comp} is a case study which shows how phoneme-level duration is affected by different model.

\begin{table}[h]
\caption{Comparison of alignment quality between different configurations. Sine-Cosine or Gaussian encoding function, trainable or fixed position encoding and CNN or UFANS decoder are evaluated based on their average difference between their duration prediction and real duration length on 100 audios.}
\centering
\begin{adjustbox}{width=0.45\textwidth}
\begin{tabular}{l|c|c|c}

\hline
Encoding func & Trainable? &  Decoder  & Average-diff  \\
\hline
Gaussian & Trainable  & CNN & 2.58 \\

Sin-Cos & Fixed & CNN & 1.96 \\ 

Sin-Cos & Trainable & UFANS & 1.80 \\

\textbf{Sin-Cos} & \textbf{Trainable} & \textbf{CNN}  & \textbf{0.85} \\
\hline
\end{tabular}
\end{adjustbox}
\label{tab : ave diff}
\end{table}

\hspace{4pt}

\begin{figure}[h]
    \centering
    \includegraphics[width=3.2in]{4824-align.pdf}
    \caption{Attention plot of text : This is the destination for all things related to development at stack overflow.\\
           Phoneme : DH IH S IH Z DH AH D EH S T AH N EY SH AH N F AO R AO L TH IH NG Z R IH L EY T IH D T UW D IH V EH L AH P M AH N T AE T S T AE K OW V ER F L OW .}
\label{fig : atten_plot}
\end{figure}

\subsubsection{Position Encoding Function and Alignment Quality}
We replace the Sine and Cosine position encoding function with Gaussian function. As Table~\ref{tab : ave diff} shows, the experimental results show that the model can not learn correct alignment with Gaussian function. We give a theoretical analysis in Appendix.

\subsubsection{Trainable Position Encoding and Alignment Quality}
We replace the trainable position encoding with a fixed position encoding. The experimental results show that the model can learn better alignment with trainable position encoding.

\subsubsection{Decoder and Alignment Quality}
In order to identify the relationship between decoder and alignment quality in stage 1, we replace simple convolutional decoder by UFANS with 6 down-sampling layers. Experiments show the computed attention width is much worse than that with the simple convolutional decoder. And the synthesized audios also suffer from error modes like repeated words and skipped words. The results show the simple decoder may be better in alignment learning stage. More details are shown in Table~\ref{tab : ave diff}.
With UFANS decoder, our model can get comparable loss no matter that the alignment is accurate or not. Therefore, alignment isn't well trained with UFANS decoder. Human is sensitive to phoneme speed, so speech will be terrible if duration in inaccurate. To solve the problem, we train the alignment with simple CNNs, then fix the alignment structure. With the fixed alignment and UFANS decoder, our model can generate high quality audio in a parallel way.

\section{Related Works}
FastSpeech \cite{DBLP:journals/corr/abs-1905-09263},which is proposed in same period, can also generate acoustic features in a parallel way. Specifically, it extract attention alignments from an auto-regressive encoder-decoder based teacher model for phoneme duration prediction, which is used by a length regulator to expand the source phoneme sequence to match the length of the target mel-spectrogram sequence for parallel mel-spectrogram generation. Using the phoneme duration extracted from an teacher model is a creative work to solve the problem that model can't inference in a parallel way. However, it's speed is still not fast enough to satisfy industrial application, especially it can't speed up when batch size is increased.

Our FPETS has lower time latency and faster than FastSpeech. On average FPETS generates 10ms per sentence under GTX 1080ti GPU and FastsSpeech is 25ms per sentence under Tesla V100 GPU, which is known faster than GTX 1080ti. And FPETS can also automatically specify the phoneme duration by trainable position encoding.

\section{Conclusion}

In this paper, a new non-autoregressive, fully parallel end-to-end TTS system, FPETS, is proposed. Given input phonemes, FPETS can predict all acoustic frames simultaneously rather than autoregressively. Specifically FPETS utilize a recent proposed U-shaped convolutional structure, which can be fully parallel  and has stronger representation ability. The fully parallel alignment structure inference alignment relationship between all phonemes and audio frames at once. The novel trainable position encoding method can utilize position information better and two-step training strategy improves the alignment quality.

FPETS can utilize the power of parallel computation and reach a significant speed up of inference compared with state-of-the-art end-to-end TTS systems. More specifically, FPETS is 600X faster than Tacotron2, 50X faster than DCTTS and 10X faster than Deep Voice3. And FPETS can generates audios with equal or better quality and fewer errors comparing with other system. As far as we know, FPETS is the first end-to-end TTS system which is fully parallel.

\bibliography{AAAI-MaD.4824.bib}
\bibliographystyle{aaai}

\end{document}